\documentclass[manuscript]{aastex}

\usepackage{aas_macros}
\usepackage{natbib}

\shortauthors{Testi et al.}

\begin{document}

\title{Hunting for planets in the HL Tau disk}
\author{L. Testi\altaffilmark{1,2}}
\affil{ESO, Karl Schwarzschild str. 2, D-85748 Garching bei Muenchen, Germany}
\email{ltesti@eso.org}

\author{A. Skemer}
\affil{Steward Observatory, University of Arizona, 933 N. Cherry Ave., Tucson, AZ 85721, USA}
\affil{University of California, Santa Cruz, 1156 High Street, Santa Cruz, CA 95064, USA}

\author{Th. Henning}
\affil{Max Planck Institute for Astronomie, K\"onigstuhl 17, D-69117 Heidelberg, Germany}

\author{V. Bailey, D. Defr\`ere, Ph. Hinz, J. Leisenring, A. Vaz}
\affil{Steward Observatory, University of Arizona, 933 N. Cherry Ave., Tucson, AZ 85721, USA}

\author{S. Esposito}
\affil{INAF-Osservatorio Astrofisico di Arcetri, Largo E. Fermi 5, I-50125 Firenze, Italy}

\author{A. Fontana}
\affil{INAF-Osservatorio Astronomico di Roma, Monte Porzio, Italy}

\author{A. Marconi}
\affil{Universit\'a degli Studi di Firenze, Dipartimento di Fisica e Astronomia, Firenze, Italy}

\author{M. Skrutskie}
\affil{University of Virginia, 530 McCormick Road, Charlottesville, VA 22904, USA}

\author{C. Veillet}
\affil{LBT Observatory, University of Arizona, 933 N. Cherry Ave., Tucson, AZ 85721, USA}

\altaffiltext{1}{INAF-Osservatorio Astrofisico di Arcetri, Largo E. Fermi 5, I-50125 Firenze, Italy}
\altaffiltext{2}{Excellence Cluster `Universe', Boltzmann str. 2, D-85748 Garching bei Muenchen, Germany}

\begin{abstract}
   Recent ALMA images of HL Tau show gaps in the dusty disk that may be caused by planetary bodies.
   Given the young age of this system, if confirmed, this finding would imply very short timescales for planet formation, probably in a gravitationally unstable disk.
   To test this scenario, we searched for young planets by means of direct imaging in the L$^\prime$-band using the Large Binocular Telescope Interferometer mid-infrared camera.
   At the location of two prominent dips in the dust distribution at $\sim$70~AU ($\sim$0.$\!^{\prime\prime}$5) from the central star we reach a contrast level of $\sim 7.5$~mag.
   We did not detect any point source at the location of the rings. Using evolutionary models we derive
   upper limits of $\sim$10-15~M$_{Jup}$ at $\le 0.5$-1~Ma for the possible planets. With these sensitivity limits we should have been able to detect companions sufficiently massive to open full gaps in the disk.
   The structures detected at mm-wavelengths 
   could be gaps in the distributions of large grains on the disk midplane, caused by planets not massive enough to fully open gaps. Future ALMA observations of the molecular gas 
   density profile and kinematics as well as higher contrast infrared observations may be able to provide a definitive answer.
\end{abstract}

\maketitle
%

\section{Introduction}
\label{s_intro}
   

   The protoplanetary disk surrounding the HL~Tau young stellar object 
   has been extensively studied in the past and was the first to be kinematically
   resolved \citep[e.g.]{1991ApJ...382L..31S}. The system is known to be
   very young with an envelope still accreting onto the disk \citep[e.g.]{1993ApJ...418L..71H,1999ApJ...519..257M}. The age has been estimated to be $\le 1$~Ma, and values as low as 0.2~Ma have been claimed in the literature \citep{2011A&A...529A.105G}.
   With an estimated stellar mass of $\sim$0.7~M$_\odot$ \citep{1995ApJS..101..117K,1997ApJ...478..766C} 
   and a total disk mass in the range $\sim$0.1-0.15~M$_\odot$ 
   \citep{2011A&A...529A.105G,2011ApJ...741....3K}, the disk may be very close to being 
   gravitationally unstable
   
   The latest ALMA long baselines science verification results show a series of eccentric bright and dark rings \citep{2015arXiv150302649P}, which could be consistent with the presence of young protoplanets in the disk. The pair of dark rings named D5 and D6, some of the lowest optical depth ones in the ALMA image, have radii of $\sim$65 and $\sim$75~AU ($\sim$0.$\!^{\prime\prime}$5) from the central star. These fall in the region of the disk predicted to be gravitationally unstable by \citep{2011ApJ...741....3K}. If a planetary body is indeed responsible for opening the D5/D6 gaps, given the distance from the central star, disk properties and system age, the most likely formation route for such planets would be through gravitational instabilities \citep{2014prpl.conf..643H}. The possibility of the presence of planets formed
   by gravitational instabilities can thus be tested directly with large telescopes equipped with adaptive optics and thermal infrared diffraction limited cameras. 
  
  We used the Large Binocular Telescope Interferometer mid-infrared camera (LBTI/LMIRcam) to search for the presence of young giant planets
  within the disk of HL~Tauri, specifically, but not exclusively at the location of the D5/D6 dark rings in the ALMA image. We describe the observations in Sec.~\ref{s_obs}, the results are presented and compared with model predictions in Sec.~\ref{s_res}, our conclusions are summarized in Sec.~\ref{s_con}.
  
   HL~Tauri is located in the Taurus molecular cloud and throughout this paper we will 
   adopt the commonly used distance of 140~pc \citep{2008hsf1.book..405K}.


\section{Observations}
\label{s_obs}

We observed HL~Tau with the LBTI \citep{2014SPIE.9146E..0TH} and its 1-5$\mu$m, L/M Infrared Camera \citep[LMIRcam]{2010SPIE.7735E..3HS,2012SPIE.8446E..4FL}.  LBTI/LMIRcam is used with the LBT's dual deformable secondary adaptive optics (AO) systems \citep{2011SPIE.8149E..02E} either for interferometry, or for standard AO imaging.  For the observations described in this paper, we used only one side of the LBT's two 8.4 meter diameter primary mirrors for AO imaging.  We observed HL Tau on UT November 7, 2014 and UT November 17, 2014 with the L' (3.8 $\mu$m) filter, and on UT November 19, 2014 with the K$_s$ (2.2 $\mu$m) filter.  Seeing during these observations was measured by an on-site DIMM to be 1.0-2.0$^{\prime\prime}$, 0.$\!^{\prime\prime}$8-1.$\!^{\prime\prime}$1, and 0.$\!^{\prime\prime}$7-0.$\!^{\prime\prime}$9 respectively.  The observations comprised 70, 44 and 47 minutes of data with 59, 63 and 59 degrees of parallactic angle rotation respectively.  

HL Tau is R$\sim$14 magnitudes, which is faint for most natural guide star AO systems, but the LBTAO system is more sensitive than other AO systems because it uses (a) a pyramid wavefront sensor and (b) on-chip detector binning to reduce read out noise level when operating with faint stars \citep{2001A&A...369L...9E,2011SPIE.8149E..02E}.  Another option would be to guide on XZ Tau, a 0.$\!^{\prime\prime}$3 binary that is $\sim$30$^{\prime\prime}$ from HL Tau.  At the time of our observations, XZ Tau was $\sim$2 magnitudes brighter than HL Tau on the AO wavefront sensor.  We used HL Tau as our guide-star for the November 7 observations, and then XZ Tau for the subsequent observations.  Guiding on XZ Tau produced higher Strehl ratio images of HL Tau.  However, isoplanatic effects stretched the HL Tau image in the direction of the guide-star.

\begin{figure}
   \centering
   \includegraphics[width=9.5cm]{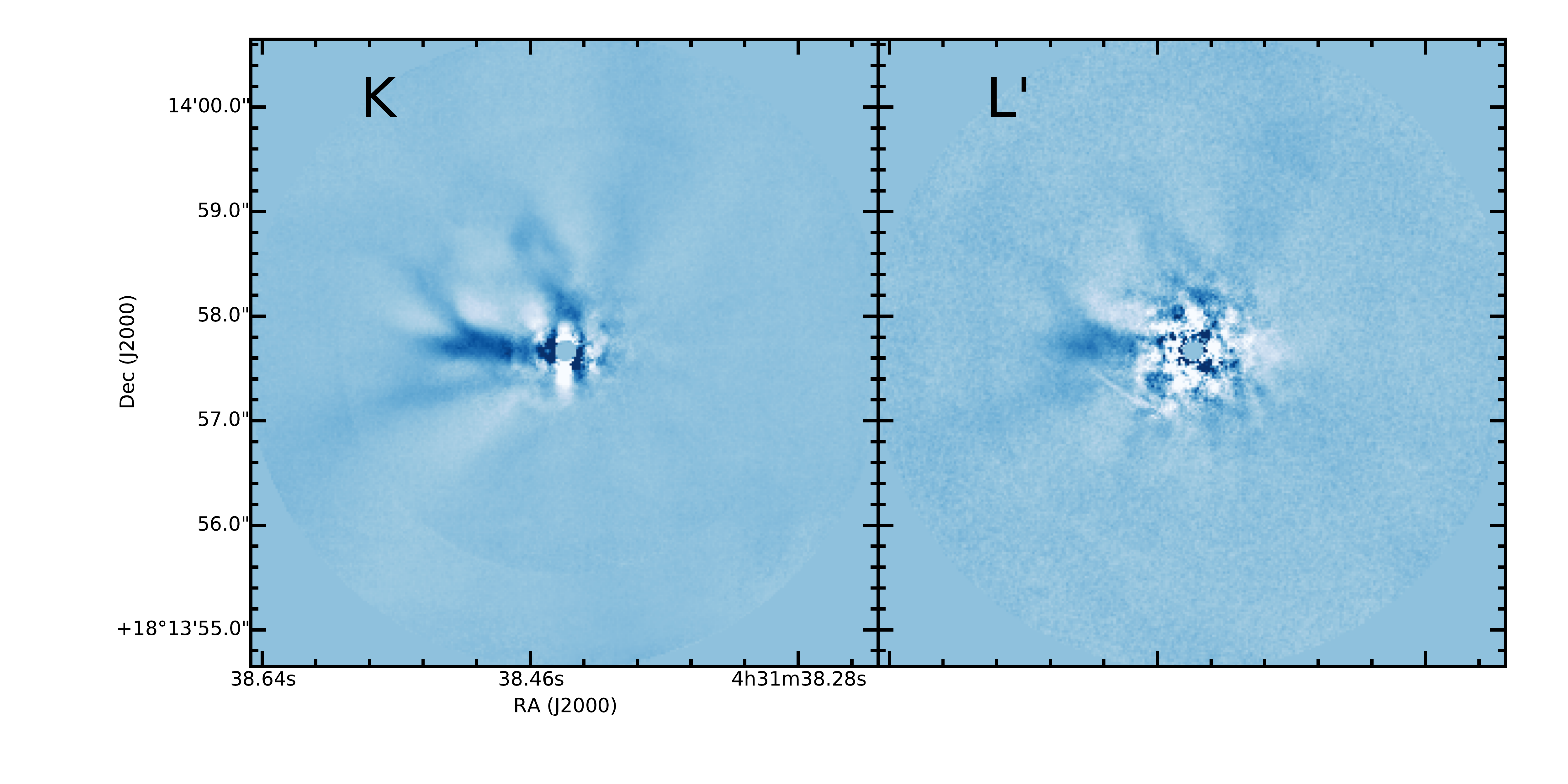}
      \caption{
      LBTI K$_s$ and L$^{\prime}$ images of the HL~Tau system (the star is hidden by the central mask). Scattered light emission from the disk and envelope is visible. The stellar PSF has been modeled and subtracted.
              }
         \label{f_lbt_obs}
  \end{figure}

We processed the images with the high-contrast optimization pipeline developed for the LBTI Exozodi Exoplanet Common Hunt survey \citep[LEECH; ][]{2014SPIE.9148E..0LS,2015A&A...576A.133M}.  After basic infrared processing (nod-subtracting, registering, coadding), we model and subtract the star using angular differential imaging \citep[ADI; ][]{2006ApJ...641..556M} and principle component analysis \citep[PCA; ][]{2012ApJ...755L..28S,2012MNRAS.427..948A}.  Our final Ks and L' images are shown in Figure~\ref{f_lbt_obs}.
There is a clear fixed residual between the two images, which is scattered light from HL Tau's envelope, as processed by PCA, which self-subtracts (hence the negative residuals).

The structure of the scattered light envelope impacts our ability to search for point-source exoplanet companions.  However, cool exoplanets are much redder than scattered starlight at these wavelengths \citep[Ks-L'$\sim$2 mag for COND/DUSTY models at our detection threshold; ][]{2003A&A...402..701B,2000ApJ...542..464C}, so a putative exoplanet would appear as a point source in the L' image, while being much fainter or undetectable in the Ks image.  To search for red companions, we subtract the Ks image from the L' image\footnote{This is analogous to simultaneous differential imaging \citep{1999PASP..111..587R}.}, scaled so that the scattered light envelope is suppressed as much as possible.  
We optimize the overall scaling and the number of principle components used to subtract the stellar PSF at every radius, as described in detail in \citet{2015A&A...576A.133M} and briefly described in the following.  Artificial point sources are inserted into the original data, one-by-one, at 8 different azimuths per radius.  For each artificial point source we run our PCA star subtraction code, iteratively adjusting the brightness of the artificial planet until it reaches 5 times the standard deviation of a diffraction limit size smoothed annulus at the same radius.  The number of principle components that gives the faintest 5-sigma artificial planets is used to produce the final L' image (Figure~\ref{f_image}).  The average magnitude of the 5-sigma artificial planets at each radius is plotted as our contrast curve (Figure~\ref{f_contrast}).

\section{Results}
\label{s_res}

\begin{figure}
   \centering
   \includegraphics[width=7.5cm]{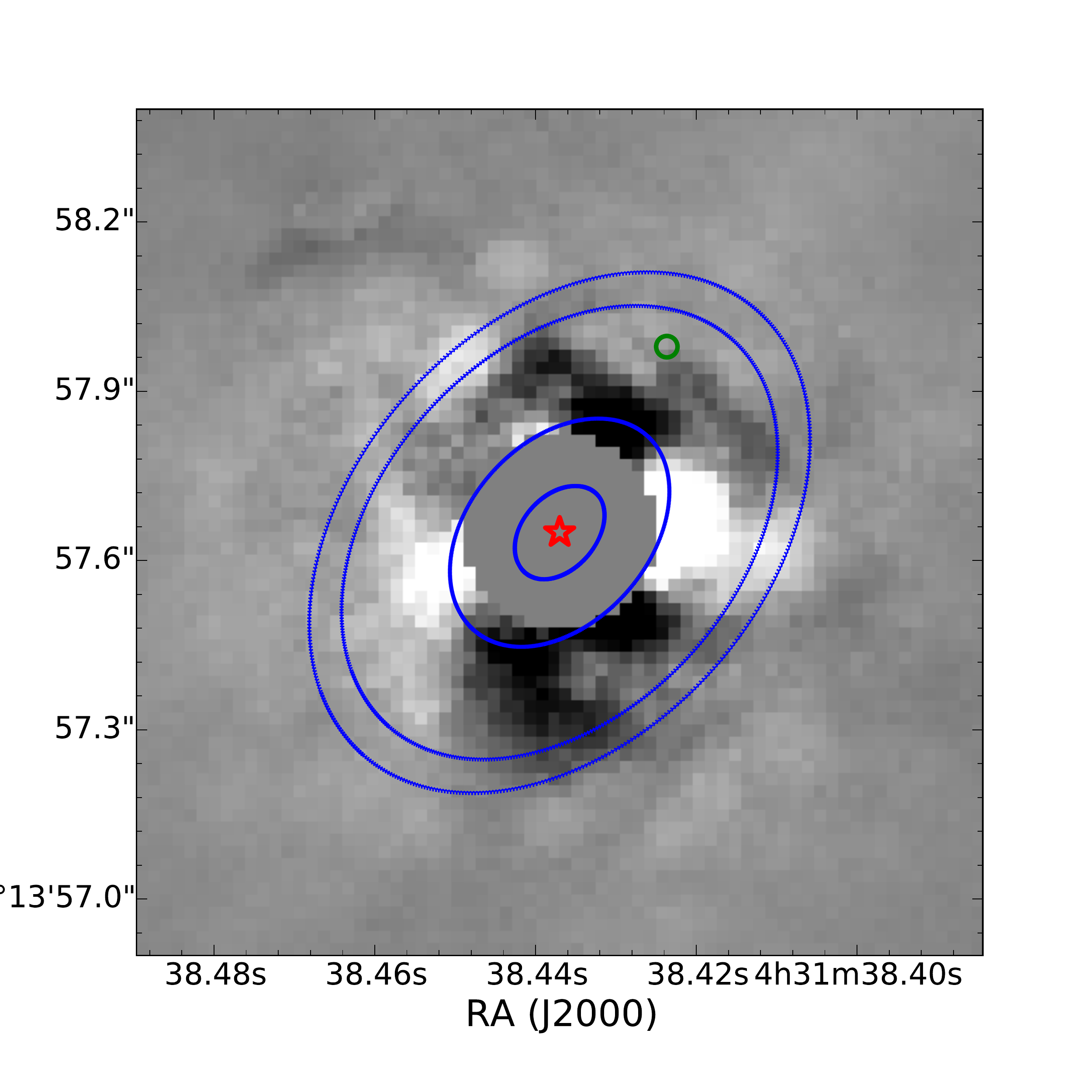}
      \caption{
      LBTI L$^{\prime}$ image of the HL~Tau system (the star is hidden by the central mask), with scattered light emission mitigated as described in the text. 
      No point source is detected in the image. The red cross marks the position of the young star. The green circle shows the position of the unconfirmed protoplanet reported by \citet{2008MNRAS.391L..74G}. The blue ellipses mark the position of the highest contrast dark rings in the ALMA image \citep[D1, D2, D5 and D6, as defined in][]{2015arXiv150302649P}.
              }
         \label{f_image}
  \end{figure}

In Figure~\ref{f_image} we show as blue ellipses the location of the highest contrast dark rings in the ALMA images: D1, D2, D5 and D6, as defined by \citet{2015arXiv150302649P}. We also show the location of HL~Tau with a red star, and the position of the unconfirmed protoplanet reported by \citet[][green circle]{2008MNRAS.391L..74G}. The inner masked region has a radius of $\sim$0.$\!^{\prime\prime}$18. We do not detect any point source in the image, however this has to be interpreted with caution as the point source sensitivity is a strong function of the distance from the central star. 

\begin{figure}
   \centering
   \includegraphics[width=7.8cm]{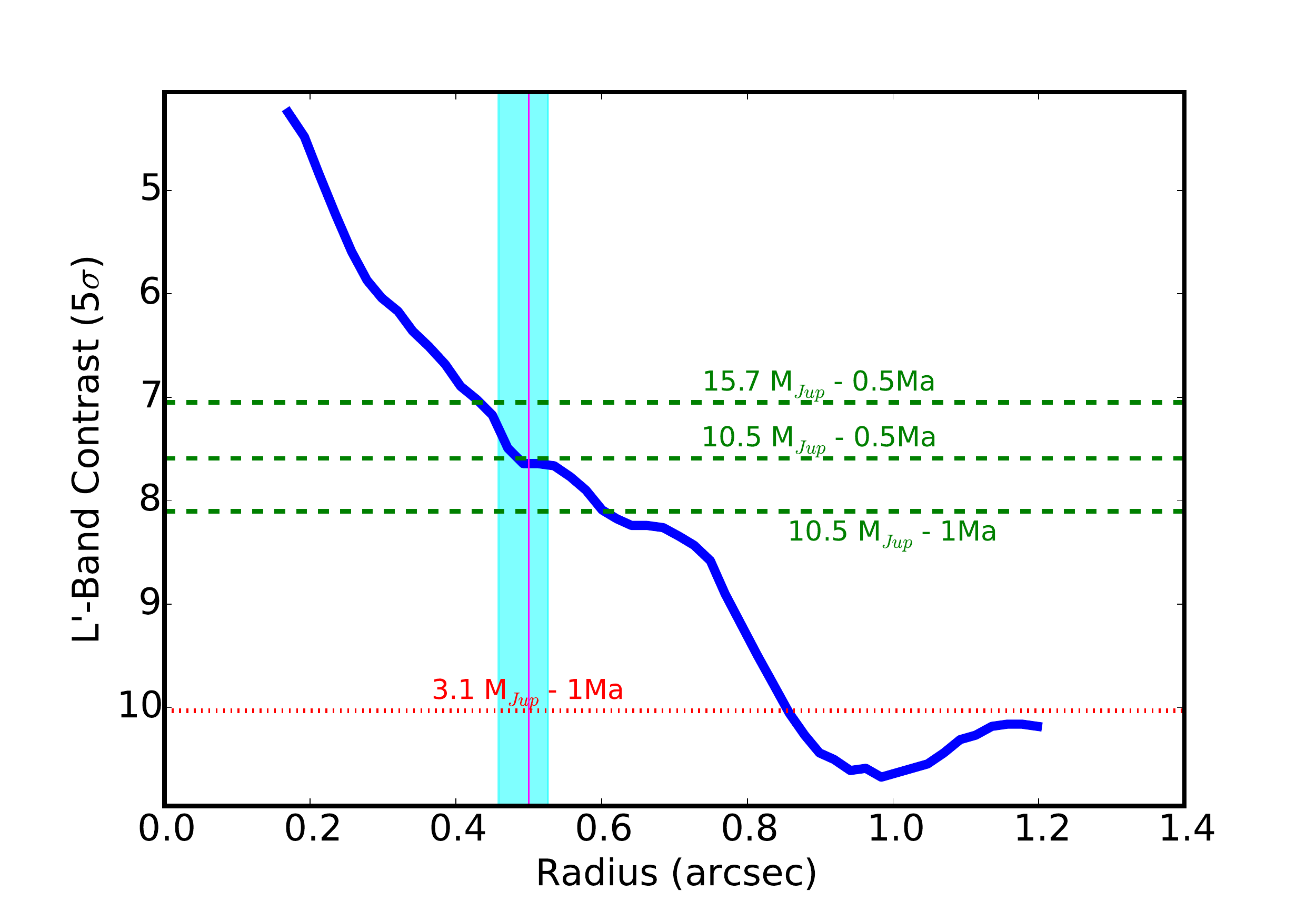}
      \caption{
      L$^\prime$ contrast as a function of angular distance from HL~Tau. The blue line shows the 5$\sigma$ detection limit for point sources computed from the image shown in Figure~\ref{f_image}. The vertical cyan band marks the approximate location of the D5/D6 rings in the ALMA image. The dotted red line shows the expected emission level for a 3.1~M$_{Jup}$ planet, assuming an age of 1~Ma and the evolutionary tracks of \citet{2008A&A...482..315B}. The three green dashed lines show the expected emission for 10.5~M$_{Jup}$ at 0.5 and 1~Ma and $\sim$15.7~M$_{Jup}$ at 0.5~Ma, as labeled, using the evolutionary tracks of \citet{2015arXiv150304107B}. See text for details.
              }
         \label{f_contrast}
  \end{figure}

In Figure~\ref{f_contrast} we show the computed 5$\sigma$ contrast limit for a point source detection in the image of Figure~\ref{f_image} (see Sect.~\ref{s_obs}). The minimum contrasts for detection are approximately $\sim$4.5 magnitudes at the location of D2 ($\sim$0.$\!^{\prime\prime}$2), $\sim$7.5 at the location of D5 and D6 ($\sim$0.$\!^{\prime\prime}$5), and exceeding 10 beyond $\sim$0.$\!^{\prime\prime}$85 ($\sim$120~AU). At the location of the candidate protoplanet reported by \citet{2008MNRAS.391L..74G}, the contrast limit for detection is $\sim$7. To convert these contrast levels in an absolute magnitude limit we need an estimate of the L$^{\prime}$ magnitude of the star. Seeing limited L-band photometry is available \citep[e.g.][]{1995ApJS..101..117K}, but the source has a significant extended 
component that is well resolved in our high angular resolution images.

\citet{1997ApJ...478..766C} estimated that the compact component of the emission in their 0.$\!^{\prime\prime}$2 resolution images is approximately 
0.6 magnitudes fainter than the total emission at K$^\prime$. Assuming a similar ratio for the L$^\prime$ band and using the (K-L) color from 
\citet{1995ApJS..101..117K}, we estimate an absolute magnitude of L$\sim$0.5
for the central point source.  
To obtain the intrinsic absolute magnitudes, we additionally need an estimate of the extinction. The extinction towards HL~Tau has been quoted in the range 7-24 in the V band \citep[e.g.][]{1997ApJ...478..766C}. Using the extinction law of \citep{1985ApJ...288..618R} this corresponds to 0.4-1.3 magnitudes in L$^\prime$. Here we will use a value of $\sim$1, which need to be subtracted to the limits derived above. Note that in this procedure we are neglecting the possible extinction provided by the disk material, which implies that the limits we are deriving strictly apply only to planets that have completely cleared a gap in the disk dust distribution. 

Our final estimate of the L' intrinsic absolute magnitude for the unresolved component in the HL~Tau system is thus $\sim -0.5$~mag. We will use this value to estimate the absolute magnitudes that correspond to our sensitivity limits, by adding the value to the contrast plotted in Figure~\ref{f_contrast}. The three sources of uncertainty on this estimate are the fraction of extended emission, the exact extinction value, and the possible photometric variability of the central star. To estimate the latter, we checked the WISE data base and found a peak-to-peak variation of $\sim 0.4$~mag\footnote{The WISE photometry is uncertain as HL~Tau is in the saturated sources regime; the source has measured magnitudes in the range 5.0-5.4 in the WISE W1 band over a period of 200 days}.
Considering all these factors, we estimate an overall uncertainty of $\sim$0.5~mag.

To convert the absolute magnitudes in an estimate of the planetary companion masses, we used the evolutionary tracks of \citet{2008A&A...482..315B,2015arXiv150304107B}. As discussed by several authors, the evolutionary tracks are very uncertain at young ages, as the 
initial conditions of protoplanets may depend on the formation scenario \citep[see the discussion in][]{2010RPPh...73a6901B}. Planets forming via core accretion may be expected to have cooler initial conditions, leading to smaller radii and fainter magnitudes for the same mass as compared to the canonical evolutionary tracks \citep{2007ApJ...655..541M}. The recent analysis by \citet{2014MNRAS.437.1378M} has shown that few directly imaged exoplanets at large radii have physical parameters consistent with warm initial conditions. Their results rule out cold initial conditions for the exoplanets in the HR~8799, 2MASS J12073346$-$3932539, and $\beta$~Pictoris systems. Warm initial conditions are thus confirmed for giant planets at large distances from the central star, and potentially also for planets formed via core accretion (as it may be the case for $\beta$~Pic~b). In our case, given the young age of the system, the fact that the disk appears to be marginally unstable, and the large distances from the star we are probing with our observations, it is most likely that any large planet would have been formed via gravitational instability, rather than core accretion. We thus expect that the upper limits on the planetary masses derived from evolutionary tracks with warm initial conditions will be the most appropriate. The earliest ages for which the tracks are tabulated is 1~Ma in \citet{2008A&A...482..315B} and 0.5~Ma in \citet[][note that these authors have $\sim$10~M$_{Jup}$ as the lowest mass for their computations]{2015arXiv150304107B}. 
As the system is believed to be younger than 0.5-1~Ma, and the possible companions even younger, it is likely that the 5$\sigma$ upper limits that we derive are  conservative.

\begin{figure}
   \centering
   \includegraphics[width=7.8cm]{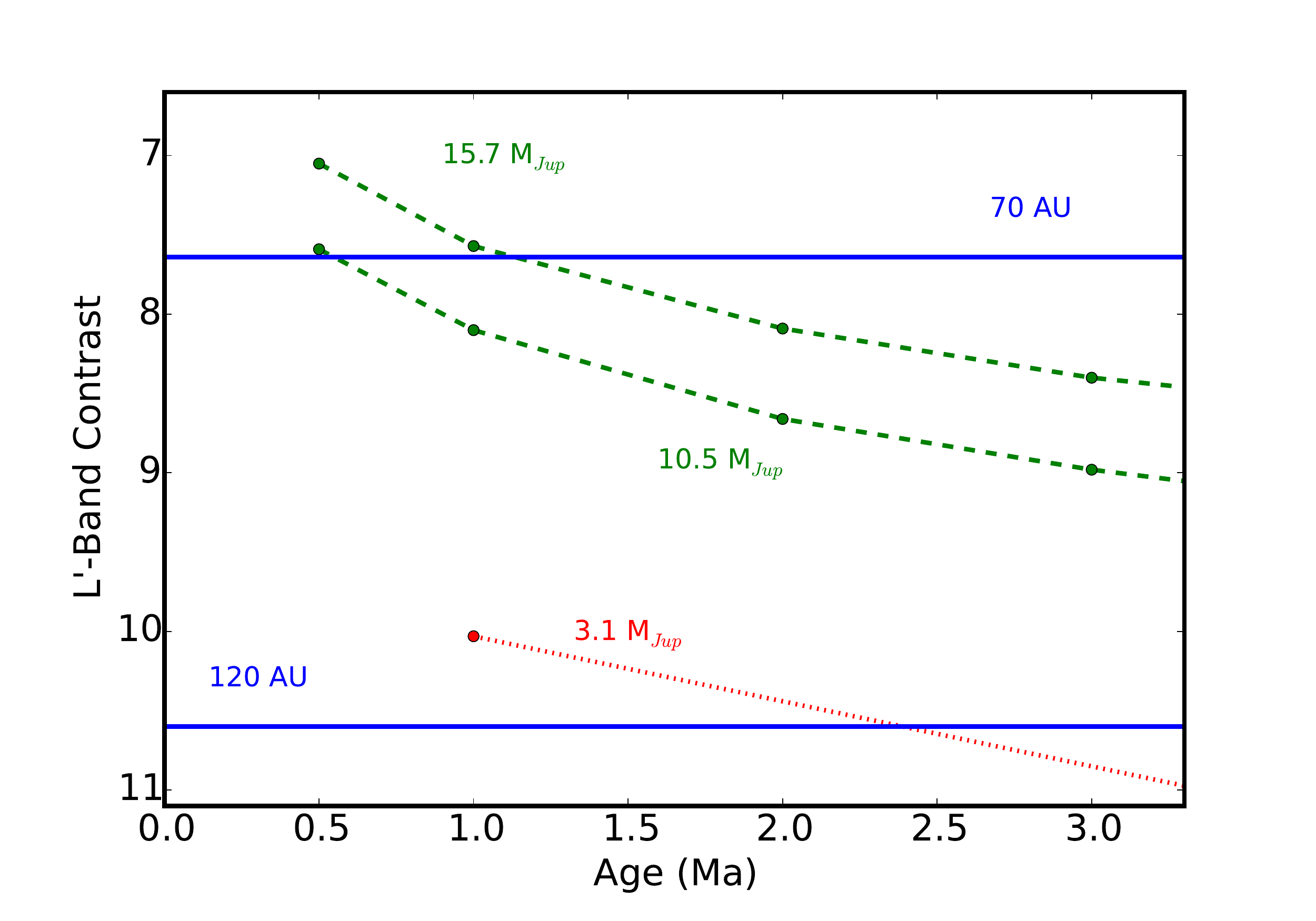}
      \caption{
      Contrast as a function of age for young planets with masses 3.1~M$_{Jup}$ from \citet[][red circles and dotted lines]{2008A&A...482..315B}, and 10.5 and 15.7~M$_{Jup}$ from \citet[][green circles and dashed lines]{2015arXiv150304107B}, with the assumptions discussed in the text.  Blue horizontal lines show the 5$\sigma$ detection limits for point sources at 70 and 120~AU as derived from Figure~\ref{f_contrast}. 
              }
         \label{f_c_age}
  \end{figure}

We report the results of our analysis in Figures~\ref{f_contrast} and~\ref{f_c_age}. In Figure~\ref{f_contrast} we show the contrast predicted by evolutionary tracks for four different set of parameters: planets with masses $\sim$3 and $\sim$16~M$_{Jup}$ at 1~Ma age from \citet{2008A&A...482..315B} and $\sim$16~M$_{Jup}$ at 0.5~Ma and $\sim 10.5$~M$_{Jup}$ at 0.5 and 1~Ma from \citet{2015arXiv150304107B}. Given the expected age of the system, and considering a 0.5~mag uncertainty in the absolute magnitude value due to the uncertainties discussed above, planets more massive than $\sim$10-15~M$_{Jup}$ should be detected if present at the location of the D5/D6 rings and at larger distances in the disks. In Figure~\ref{f_c_age} we show how planetary evolutionary tracks for 10 and~15.7~M$_{Jup}$ compare with our 5$\sigma$ detection limits at 70~AU. For completeness, we also show a track for 3.1~M$_{Jup}$ and the contrast limit at 120~AU, suggesting that it would be detectable in the outer disk. Nevertheless, such a planet would most likely be unable to clear a full gap in the disk and would suffer a higher extinction than our estimate (see below). 

\section{Discussion}
\label{s_dis}

\begin{figure}
   \centering
   \includegraphics[width=7.8cm]{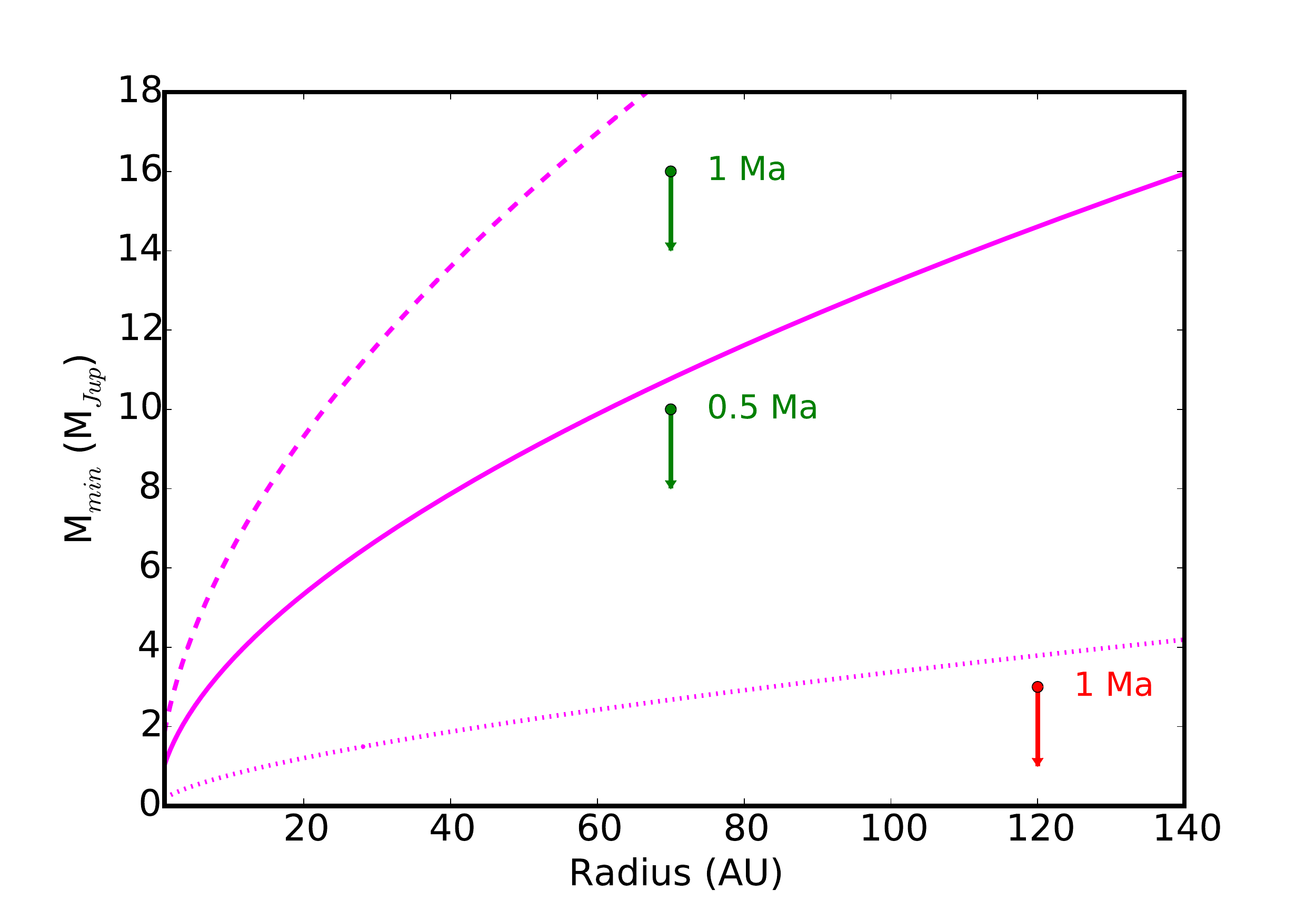}
      \caption{
      Gap opening criterion for the fiducial disk model discussed in the text, derived using the formula in \citet{2014prpl.conf..667B}, shown as a full magenta line for the reference viscosity value ($\alpha = 0.1$). Dotted and dashed lines show the effect of reducing ($\alpha = 0.001$) or increasing ($\alpha = 0.2$) the viscosity. The 5$\sigma$ upper limits for 0.5 and 1~Ma at 70~AU derived from our observations are shown as green  symbols, respectively. For completeness we also show our limit for 1~Ma at 120~AU (red symbol). See text for details.
              }
         \label{f_gap_open}
  \end{figure}

Using a fiducial disk model around a $\sim$1~M$_\odot$ mass star, with a temperature profile on the midplane $T(r)=$300~K$\times (r/1~{\rm AU})^{0.5}$ and a viscosity parameter $\alpha$=0.01, we can use the formula provided in \citet{2014prpl.conf..667B} for the gap opening criterion derived by \citet{2006Icar..181..587C} to estimate the minimum mass for a planet to carve a gap in the disk. The comparison of these results with our planetary mass upper limits are shown in Figure~\ref{f_gap_open}. At $\sim$70~AU from the star, our upper limits are comparable to the minimum mass derived using the \citet{2006Icar..181..587C} criterion, depending on the exact age of the planets. In our analysis we have assumed an age range of 0.5-1.0~Ma for the possible planets (based on the estimated age for the star), but these would most certainly be upper limits to the age of planets. Our estimated planetary mass upper limits are thus conservative, as younger planets would be brighter. \citet{2015ApJ...802...56M} have recently examined the gap opening effectiveness of giant planets undergoing migration in disks, and found that planets significantly above the \citet{2006Icar..181..587C} criterion may still be unable to open a gap. In particular, a 15~M$_{Jup}$ planet in all their simulations were unable to open a gap in the outer disk. In addition, we note that the planetary mass required to open a gap in the disk is a very strong function of the disk viscosity, for gravitationally unstable disks larger values of $\alpha$ may be appropriate. 


We do not detect planets at the location of the D5/D6 gaps detected by ALMA, down to the mass limits that would be expected for companions with the ability to fully clear a gap. Nevertheless, there are still possible alternatives, that involve the presence of planets, that are still plausible for the HL~Tau disk. The ALMA observations reported by  \citet{2015arXiv150302649P} are mostly sensitive to the distribution of the millimetre size grains. \citet{2014ApJ...785..122Z} have shown that even very small planetary cores are capable of creating disturbances in the gas distribution that are sufficient to efficiently trap the large grains. \citet{2015MNRAS.453L..73D} analyzed in more detail the case of HL~Tau and showed that planets as small as $\sim 0.2-0.5$~M$_{Jup}$ may be able to explain the ALMA observations. Relatively small giant planets, smaller than our detection limit, could thus be present in the disk and produce the effects observed by ALMA, if the disk viscosity is very low. 
For disk parameters very similar to those of HL~Tau, \citet{2011ApJ...731...74B} has shown that $\sim$5~M$_{Jup}$ planets could form at 70~AU through disk gravitational instabilities. Such planets, if formed very early in the disk and if they have not migrated significantly in the following 0.5-1~Ma, may have escaped our detection and, while being unable to carve proper gaps in the gaseous disk, would have a sufficient mass to produce the observed confinement of the millimeter size grains.

\section{Conclusions}
\label{s_con}

We have searched for possible young giant planets in the HL~Tau disk using the LBTI. We do not detect point sources within the disk, in the parameter range probed by our observations. More specifically, no companions with mass above $\sim$10-15~M$_{Jup}$ and age $\le 0.5-1$~Ma is detected at the location of the 
gap detected by ALMA in the dust emission at $\sim$70~AU from the star, under the assumption that such planet would have mostly cleared a gap in the vertical disk dust distribution.

Our limits imply that large planets capable of opening a full gap in the gaseous disk at $\sim$70~AU cannot be the explanation for the observed distribution of dust grains as observed with ALMA. If planets  are present in the HL~Tau disk they are relatively small giant planets that have not opened a full gap in the disk. This conclusion is consistent with recent theoretical analysis \citep[e.g.][]{2015MNRAS.453L..73D}.


There are two obvious observational avenues to further constraint the origin of the large grains confinement in HL~Tau: constrain the gas distribution and properties (e.g. viscosity) and improve on our planet detection limits. The former can be attempted with ALMA, although the results of the Science Verification observations show that it will require to find tracers less affected by the envelope emission and a substantial investment of observing time. To completely exclude the presence of giant planets formed by gravitational instabilities at the location of the D5/D6 dark rings, will require an improvement of $\sim$3-4~mag in the contrast at $\sim$0.$\!^{\prime\prime}$5 from the central star.

\acknowledgments{
We thank the LBT Observatory director and staff for the flexibility in scheduling this project and for supporting the observations. This publication makes use of data products from the Wide-field Infrared Survey Explorer, which is a joint project of the University of California, Los Angeles, and the Jet Propulsion Laboratory/California Institute of Technology, funded by the National Aeronautics and Space Administration. The LBT is an international collaboration among institutions in the United States, Italy and Germany. LBT Corporation partners are: The University of Arizona on behalf of the Arizona university system; Istituto Nazionale di Astrofisica, Italy; LBT Beteiligungsgesellschaft, Germany, representing the Max-Planck Society, the Astrophysical Institute Potsdam, and Heidelberg Univer- sity; The Ohio State University, and The Research Corporation, on behalf of The University of Notre Dame, University of Minnesota and University of Virginia.
A.S is supported by the National Aeronautics and Space Administration through Hubble Fellowship grant HSTHF2-51349 awarded by the Space Telescope Science Institute, which is operated by the Association of Universities for Research in Astronomy, Inc., for NASA, under contract NAS 5-26555.  Development of the LEECH high-contrast imaging pipeline is funded by the NASA Origins of Solar Systems Program, grant NNX13AJ17G.  The Large Binocular Telescope Interferometer is funded by the National Aeronautics and Space Administration as part of its Exoplanet Exploration program.  LMIRcam is funded by the National Science Foundation through grant NSF AST-0705296.
This work was partly supported by the Italian Ministero dell\'\,Istruzione, Universit\`a e Ricerca through the grant Progetti Premiali 2012 -- iALMA (CUP C52I13000140001).
}


\begin{thebibliography}{}
\expandafter\ifx\csname natexlab\endcsname\relax\def\natexlab#1{#1}\fi

\bibitem[{{ALMA Partnership} {et~al.}(2015){ALMA Partnership}, {Brogan},
  {Perez}, {Hunter}, {Dent}, {Hales}, {Hills}, {Corder}, {Fomalont},
  {Vlahakis}, {Asaki}, {Barkats}, {Hirota}, {Hodge}, {Impellizzeri}, {Kneissl},
  {Liuzzo}, {Lucas}, {Marcelino}, {Matsushita}, {Nakanishi}, {Phillips},
  {Richards}, {Toledo}, {Aladro}, {Broguiere}, {Cortes}, {Cortes}, {Dhawan},
  {Espada}, {Galarza}, {Garcia-Appadoo}, {Guzman-Ramirez}, {Humphreys}, {Jung},
  {Kameno}, {Laing}, {Leon}, {Marconi}, {Nikolic}, {Nyman}, {Radiszcz},
  {Remijan}, {Rodon}, {Sawada}, {Takahashi}, {Tilanus}, {Vila Vilaro},
  {Watson}, {Wiklind}, {Akiyama}, {Chapillon}, {de Gregorio}, {Di Francesco},
  {Gueth}, {Kawamura}, {Lee}, {Nguyen Luong}, {Mangum}, {Pietu}, {Sanhueza},
  {Saigo}, {Takakuwa}, {Ubach}, {van Kempen}, {Wootten}, {Castro-Carrizo},
  {Francke}, {Gallardo}, {Garcia}, {Gonzalez}, {Hill}, {Kaminski}, {Kurono},
  {Liu}, {Lopez}, {Morales}, {Plarre}, {Schieven}, {Testi}, {Videla},
  {Villard}, {Andreani}, {Hibbard}, \& {Tatematsu}}]{2015arXiv150302649P}
{ALMA Partnership}, {Brogan}, C.~L., {Perez}, L.~M., {et~al.} 2015, ArXiv
  e-prints, arXiv:1503.02649

\bibitem[{{Amara} \& {Quanz}(2012)}]{2012MNRAS.427..948A}
{Amara}, A., \& {Quanz}, S.~P. 2012, \mnras, 427, 948

\bibitem[{{Baraffe} {et~al.}(2008){Baraffe}, {Chabrier}, \&
  {Barman}}]{2008A&A...482..315B}
{Baraffe}, I., {Chabrier}, G., \& {Barman}, T. 2008, \aap, 482, 315

\bibitem[{{Baraffe} {et~al.}(2010){Baraffe}, {Chabrier}, \&
  {Barman}}]{2010RPPh...73a6901B}
---. 2010, Reports on Progress in Physics, 73, 016901

\bibitem[{{Baraffe} {et~al.}(2003){Baraffe}, {Chabrier}, {Barman}, {Allard}, \&
  {Hauschildt}}]{2003A&A...402..701B}
{Baraffe}, I., {Chabrier}, G., {Barman}, T.~S., {Allard}, F., \& {Hauschildt},
  P.~H. 2003, \aap, 402, 701

\bibitem[{{Baraffe} {et~al.}(2015){Baraffe}, {Homeier}, {Allard}, \&
  {Chabrier}}]{2015arXiv150304107B}
{Baraffe}, I., {Homeier}, D., {Allard}, F., \& {Chabrier}, G. 2015, ArXiv
  e-prints, arXiv:1503.04107

\bibitem[{{Baruteau} {et~al.}(2014){Baruteau}, {Crida}, {Paardekooper},
  {Masset}, {Guilet}, {Bitsch}, {Nelson}, {Kley}, \&
  {Papaloizou}}]{2014prpl.conf..667B}
{Baruteau}, C., {Crida}, A., {Paardekooper}, S.-J., {et~al.} 2014, Protostars
  and Planets VI, 667

\bibitem[{{Boss}(2011)}]{2011ApJ...731...74B}
{Boss}, A.~P. 2011, \apj, 731, 74

\bibitem[{{Chabrier} {et~al.}(2000){Chabrier}, {Baraffe}, {Allard}, \&
  {Hauschildt}}]{2000ApJ...542..464C}
{Chabrier}, G., {Baraffe}, I., {Allard}, F., \& {Hauschildt}, P. 2000, \apj,
  542, 464

\bibitem[{{Close} {et~al.}(1997){Close}, {Roddier}, {J.~Northcott}, {Roddier},
  \& {Elon Graves}}]{1997ApJ...478..766C}
{Close}, L.~M., {Roddier}, F., {J.~Northcott}, M., {Roddier}, C., \& {Elon
  Graves}, J. 1997, \apj, 478, 766

\bibitem[{{Crida} {et~al.}(2006){Crida}, {Morbidelli}, \&
  {Masset}}]{2006Icar..181..587C}
{Crida}, A., {Morbidelli}, A., \& {Masset}, F. 2006, \icarus, 181, 587

\bibitem[{{Dipierro} {et~al.}(2015){Dipierro}, {Price}, {Laibe}, {Hirsh},
  {Cerioli}, \& {Lodato}}]{2015MNRAS.453L..73D}
{Dipierro}, G., {Price}, D., {Laibe}, G., {et~al.} 2015, \mnras, 453, L73

\bibitem[{{Esposito} \& {Riccardi}(2001)}]{2001A&A...369L...9E}
{Esposito}, S., \& {Riccardi}, A. 2001, \aap, 369, L9

\bibitem[{{Esposito} {et~al.}(2011){Esposito}, {Riccardi}, {Pinna}, {Puglisi},
  {Quir{\'o}s-Pacheco}, {Arcidiacono}, {Xompero}, {Briguglio}, {Agapito},
  {Busoni}, {Fini}, {Argomedo}, {Gherardi}, {Brusa}, {Miller}, {Guerra},
  {Stefanini}, \& {Salinari}}]{2011SPIE.8149E..02E}
{Esposito}, S., {Riccardi}, A., {Pinna}, E., {et~al.} 2011, in Society of
  Photo-Optical Instrumentation Engineers (SPIE) Conference Series, Vol. 8149,
  Society of Photo-Optical Instrumentation Engineers (SPIE) Conference Series,
  2

\bibitem[{{Greaves} {et~al.}(2008){Greaves}, {Richards}, {Rice}, \&
  {Muxlow}}]{2008MNRAS.391L..74G}
{Greaves}, J.~S., {Richards}, A.~M.~S., {Rice}, W.~K.~M., \& {Muxlow}, T.~W.~B.
  2008, \mnras, 391, L74

\bibitem[{{Guilloteau} {et~al.}(2011){Guilloteau}, {Dutrey}, {Pi{\'e}tu}, \&
  {Boehler}}]{2011A&A...529A.105G}
{Guilloteau}, S., {Dutrey}, A., {Pi{\'e}tu}, V., \& {Boehler}, Y. 2011, \aap,
  529, A105

\bibitem[{{Hayashi} {et~al.}(1993){Hayashi}, {Ohashi}, \&
  {Miyama}}]{1993ApJ...418L..71H}
{Hayashi}, M., {Ohashi}, N., \& {Miyama}, S.~M. 1993, \apjl, 418, L71

\bibitem[{{Helled} {et~al.}(2014){Helled}, {Bodenheimer}, {Podolak}, {Boley},
  {Meru}, {Nayakshin}, {Fortney}, {Mayer}, {Alibert}, \&
  {Boss}}]{2014prpl.conf..643H}
{Helled}, R., {Bodenheimer}, P., {Podolak}, M., {et~al.} 2014, Protostars and
  Planets VI, 643

\bibitem[{{Hinz} {et~al.}(2014){Hinz}, {Bailey}, {Defr{\`e}re}, {Downey},
  {Esposito}, {Hill}, {Hoffmann}, {Leisenring}, {Montoya}, {McMahon},
  {Puglisi}, {Skemer}, {Skrutskie}, {Vaitheeswaran}, \&
  {Vaz}}]{2014SPIE.9146E..0TH}
{Hinz}, P., {Bailey}, V.~P., {Defr{\`e}re}, D., {et~al.} 2014, in Society of
  Photo-Optical Instrumentation Engineers (SPIE) Conference Series, Vol. 9146,
  Society of Photo-Optical Instrumentation Engineers (SPIE) Conference Series,
  0

\bibitem[{{Kenyon} {et~al.}(2008){Kenyon}, {G{\'o}mez}, \&
  {Whitney}}]{2008hsf1.book..405K}
{Kenyon}, S.~J., {G{\'o}mez}, M., \& {Whitney}, B.~A. 2008, {Low Mass Star
  Formation in the Taurus-Auriga Clouds}, ed. B.~{Reipurth}, 405

\bibitem[{{Kenyon} \& {Hartmann}(1995)}]{1995ApJS..101..117K}
{Kenyon}, S.~J., \& {Hartmann}, L. 1995, \apjs, 101, 117

\bibitem[{{Kwon} {et~al.}(2011){Kwon}, {Looney}, \&
  {Mundy}}]{2011ApJ...741....3K}
{Kwon}, W., {Looney}, L.~W., \& {Mundy}, L.~G. 2011, \apj, 741, 3

\bibitem[{{Leisenring} {et~al.}(2012){Leisenring}, {Skrutskie}, {Hinz},
  {Skemer}, {Bailey}, {Eisner}, {Garnavich}, {Hoffmann}, {Jones}, {Kenworthy},
  {Kuzmenko}, {Meyer}, {Nelson}, {Rodigas}, {Wilson}, \&
  {Vaitheeswaran}}]{2012SPIE.8446E..4FL}
{Leisenring}, J.~M., {Skrutskie}, M.~F., {Hinz}, P.~M., {et~al.} 2012, in
  Society of Photo-Optical Instrumentation Engineers (SPIE) Conference Series,
  Vol. 8446, Society of Photo-Optical Instrumentation Engineers (SPIE)
  Conference Series, 4

\bibitem[{{Maire} {et~al.}(2015){Maire}, {Skemer}, {Hinz}, {Desidera},
  {Esposito}, {Gratton}, {Marzari}, {Skrutskie}, {Biller}, {Defr{\`e}re},
  {Bailey}, {Leisenring}, {Apai}, {Bonnefoy}, {Brandner}, {Buenzli}, {Claudi},
  {Close}, {Crepp}, {De Rosa}, {Eisner}, {Fortney}, {Henning}, {Hofmann},
  {Kopytova}, {Males}, {Mesa}, {Morzinski}, {Oza}, {Patience}, {Pinna},
  {Rajan}, {Schertl}, {Schlieder}, {Su}, {Vaz}, {Ward-Duong}, {Weigelt}, \&
  {Woodward}}]{2015A&A...576A.133M}
{Maire}, A.-L., {Skemer}, A.~J., {Hinz}, P.~M., {et~al.} 2015, \aap, 576, A133

\bibitem[{{Malik} {et~al.}(2015){Malik}, {Meru}, {Mayer}, \&
  {Meyer}}]{2015ApJ...802...56M}
{Malik}, M., {Meru}, F., {Mayer}, L., \& {Meyer}, M. 2015, \apj, 802, 56

\bibitem[{{Marleau} \& {Cumming}(2014)}]{2014MNRAS.437.1378M}
{Marleau}, G.-D., \& {Cumming}, A. 2014, \mnras, 437, 1378

\bibitem[{{Marley} {et~al.}(2007){Marley}, {Fortney}, {Hubickyj},
  {Bodenheimer}, \& {Lissauer}}]{2007ApJ...655..541M}
{Marley}, M.~S., {Fortney}, J.~J., {Hubickyj}, O., {Bodenheimer}, P., \&
  {Lissauer}, J.~J. 2007, \apj, 655, 541

\bibitem[{{Marois} {et~al.}(2006){Marois}, {Lafreni{\`e}re}, {Doyon},
  {Macintosh}, \& {Nadeau}}]{2006ApJ...641..556M}
{Marois}, C., {Lafreni{\`e}re}, D., {Doyon}, R., {Macintosh}, B., \& {Nadeau},
  D. 2006, \apj, 641, 556

\bibitem[{{Men'shchikov} {et~al.}(1999){Men'shchikov}, {Henning}, \&
  {Fischer}}]{1999ApJ...519..257M}
{Men'shchikov}, A.~B., {Henning}, T., \& {Fischer}, O. 1999, \apj, 519, 257

\bibitem[{{Racine} {et~al.}(1999){Racine}, {Walker}, {Nadeau}, {Doyon}, \&
  {Marois}}]{1999PASP..111..587R}
{Racine}, R., {Walker}, G.~A.~H., {Nadeau}, D., {Doyon}, R., \& {Marois}, C.
  1999, \pasp, 111, 587

\bibitem[{{Rieke} \& {Lebofsky}(1985)}]{1985ApJ...288..618R}
{Rieke}, G.~H., \& {Lebofsky}, M.~J. 1985, \apj, 288, 618

\bibitem[{{Sargent} \& {Beckwith}(1991)}]{1991ApJ...382L..31S}
{Sargent}, A.~I., \& {Beckwith}, S.~V.~W. 1991, \apjl, 382, L31

\bibitem[{{Skemer} {et~al.}(2014){Skemer}, {Hinz}, {Esposito}, {Skrutskie},
  {Defr{\`e}re}, {Bailey}, {Leisenring}, {Apai}, {Biller}, {Bonnefoy},
  {Brandner}, {Buenzli}, {Close}, {Crepp}, {De Rosa}, {Desidera}, {Eisner},
  {Fortney}, {Henning}, {Hofmann}, {Kopytova}, {Maire}, {Males},
  {Millan-Gabet}, {Morzinski}, {Oza}, {Patience}, {Rajan}, {Rieke}, {Schertl},
  {Schlieder}, {Su}, {Vaz}, {Ward-Duong}, {Weigelt}, {Woodward}, \&
  {Zimmerman}}]{2014SPIE.9148E..0LS}
{Skemer}, A.~J., {Hinz}, P., {Esposito}, S., {et~al.} 2014, in Society of
  Photo-Optical Instrumentation Engineers (SPIE) Conference Series, Vol. 9148,
  Society of Photo-Optical Instrumentation Engineers (SPIE) Conference Series,
  0

\bibitem[{{Skrutskie} {et~al.}(2010){Skrutskie}, {Jones}, {Hinz}, {Garnavich},
  {Wilson}, {Nelson}, {Solheid}, {Durney}, {Hoffmann}, {Vaitheeswaran},
  {McMahon}, {Leisenring}, \& {Wong}}]{2010SPIE.7735E..3HS}
{Skrutskie}, M.~F., {Jones}, T., {Hinz}, P., {et~al.} 2010, in Society of
  Photo-Optical Instrumentation Engineers (SPIE) Conference Series, Vol. 7735,
  Society of Photo-Optical Instrumentation Engineers (SPIE) Conference Series,
  3

\bibitem[{{Soummer} {et~al.}(2012){Soummer}, {Pueyo}, \&
  {Larkin}}]{2012ApJ...755L..28S}
{Soummer}, R., {Pueyo}, L., \& {Larkin}, J. 2012, \apjl, 755, L28

\bibitem[{{Zhu} {et~al.}(2014){Zhu}, {Stone}, {Rafikov}, \&
  {Bai}}]{2014ApJ...785..122Z}
{Zhu}, Z., {Stone}, J.~M., {Rafikov}, R.~R., \& {Bai}, X.-n. 2014, \apj, 785,
  122

\end{thebibliography}
\end{document}